\documentclass[conference]{IEEEtran}
\IEEEoverridecommandlockouts

\usepackage{cite}
\usepackage{amsmath,amssymb,amsfonts}
\usepackage{textcomp}
\usepackage{xcolor}

\usepackage{booktabs}  
\usepackage{mathtools}

\usepackage{algorithm}
\usepackage{algpseudocode}

\usepackage{multirow}  
\usepackage{capt-of}  
\usepackage{mathtools}  
\usepackage[defaultlines=2,all]{nowidow}  
\usepackage{url}  
\usepackage{pgfplots}
\usepackage{tikz}
\usepackage{siunitx}

\usepackage{caption}     
\usepackage{subcaption}  

\newcommand{\basebits}{B}

\definecolor{gridcolour}{HTML}{e6e6e6}
\definecolor{colourAlgoComment}{HTML}{787878}

\definecolor{colour1}{HTML}{006ed4}  
\definecolor{colour2}{HTML}{ff7f0e}
\definecolor{colour3}{HTML}{2ca02c}
\definecolor{colour4}{HTML}{d62728}
\definecolor{colour5}{HTML}{9467bd}
\definecolor{colour6}{HTML}{8c564b}
\definecolor{colour7}{HTML}{e377c2}
\definecolor{colour8}{HTML}{7f7f7f}
\definecolor{colour9}{HTML}{bcbd22}
\definecolor{colour10}{HTML}{17becf}

\input{algorithm_customisations}
\newcommand{\algowidth}{0.94\columnwidth}
\algnewcommand{\LineComment}[1]{\textcolor{colourAlgoComment}{// \textit{#1}}}  
\newcounter{algoCounter}  
\newcommand{\showAlgoCounter}[2]{\noalign{\refstepcounter{algoCounter}#1}\textbf{Algorithm~\thealgoCounter} #2}  

\tikzset{
	cross/.style={
		cross out,
		draw,
		minimum size=2*(#1-\pgflinewidth),
		inner sep=0pt,
		outer sep=0pt},
}
\pgfdeclareplotmark{mystar}{
	\draw[white,fill=white] circle (3.3pt);
	\draw node[cross=3.0pt,very thick] {};
}

\tikzset{
	every pin/.style={
		pin edge={black,thick},
		font=\footnotesize
	},
	bits table style/.style={
		matrix of nodes,
		row sep=-\pgflinewidth,
		column sep=-\pgflinewidth,
		nodes={draw=black, font=\ttfamily},
	}
}

\newcommand{\graphStandardWidth}{0.75\columnwidth}
\newcommand{\graphStandardHeight}{0.65\columnwidth}

\pgfplotsset{
	compat=1.17,
	every axis/.style={
		font=\footnotesize,
		table/col sep=comma,
		xtick pos=bottom,
		ytick pos=left,
		tick align=outside,
		title style={yshift=-0.8em},
		legend cell align={left},
		width=\graphStandardWidth, height=\graphStandardHeight,
	},
	/pgf/declare function={Floor(\x) = round(\x-0.49);},  
	plain line plot style/.style={
		cycle list={
			{black!90, mark options={draw=black,scale=1.0}, mark=x},
			{black!90, mark options={draw=black,scale=1.0}, mark=o},
			{black!90, mark options={draw=black,scale=1.0}, mark=star},
			{black!90, mark options={draw=black,scale=1.0}, mark=square},
			{black!90, mark options={draw=black,scale=1.0}, mark=triangle},
			{black!90, mark options={draw=black,scale=1.0}, mark=diamond},
			{black!90, mark options={draw=black,scale=1.0}, mark=otimes},
			{black!90, mark options={draw=black,scale=1.0}, mark=+},
		},
		line width=0.5pt,
		legend style={
			align=left,
			column sep=0.1em,
			/tikz/every odd column/.style={yshift=0.1em},
			/tikz/nodes={inner sep=0.1em},
		},
	},
	custom line plot style/.style={
		cycle list={
			{colour1!40, mark options={draw=colour1!70, scale=1.4}, mark=*},
			{colour2!40, mark options={draw=colour2!70, scale=1.25}, mark=square*},
			{colour3!40, mark options={draw=colour3!70, scale=1.8}, mark=triangle*},
			{colour4!40, mark options={draw=colour4!70, scale=1.4}, mark=pentagon*},
			{colour5!40, mark options={draw=colour5!70, scale=1.6}, mark=diamond*},
			{colour6!40, mark options={draw=colour6!70, scale=1.4}, mark=oplus*},
			{colour7!40, mark options={draw=colour7!70, scale=1.4}, mark=10-pointed star},
			{colour8!40, mark options={draw=colour8!70, scale=1.25}, mark=halfsquare*}
		},
		line width=0.5pt,
		legend style={
			align=left,
			column sep=0.1em,
			/tikz/every odd column/.style={yshift=0.1em},
			/tikz/nodes={inner sep=0.1em},
		},
	},
	custom scatter plot style/.style={
		only marks,
		cycle list={
			{colour1, mark options={scale=1.4, line width=0.5pt, fill opacity=0.45, draw opacity=0.7}, mark=*},
			{colour2, mark options={scale=1.25, line width=0.5pt, fill opacity=0.45, draw opacity=0.7}, mark=square*},
			{colour3, mark options={scale=1.8, line width=0.5pt, fill opacity=0.45, draw opacity=0.7}, mark=triangle*},
			{colour4, mark options={scale=1.4, line width=0.5pt, fill opacity=0.45, draw opacity=0.7}, mark=pentagon*},
			{colour5, mark options={scale=1.6, line width=0.5pt, fill opacity=0.45, draw opacity=0.7}, mark=diamond*},
			{colour6, mark options={scale=1.4, line width=0.5pt, fill opacity=0.45, draw opacity=0.7}, mark=oplus*},
			{colour7, mark options={scale=1.4, line width=0.5pt, fill opacity=0.45, draw opacity=0.7}, mark=10-pointed star},
			{colour8, mark options={scale=1.25, line width=0.5pt, fill opacity=0.45, draw opacity=0.7}, mark=halfsquare*}
		},
	},
	dual boxplot style/.style={
		boxplot={
			draw direction=y,
			draw position={1/3 + Floor(\plotnumofactualtype/2) + 1/3*mod(\plotnumofactualtype,2)},
			box extend=0.3,
			every median/.style={very thick},
		},
		cycle list={
			{black, mark=*, solid, fill=colour1, fill opacity=0.25},
			{black, mark=*, solid, fill=colour2, fill opacity=0.25}
		},
		x tick label as interval,
		legend style={
			align=left,
			column sep=0.2em,
			/tikz/every odd column/.style={yshift=0.1em},
			/tikz/nodes={inner sep=0.1em},
		},
	},
	single boxplot style/.style={
		boxplot={
			draw direction=x,
			box extend=0.7,  
			every median/.style={very thick},
			mark=o,
		},
		every boxplot/.style={color=black, fill=colour1, solid, fill opacity=0.25},
	}
}

\def\BibTeX{{\rm B\kern-.05em{\sc i\kern-.025em b}\kern-.08em
    T\kern-.1667em\lower.7ex\hbox{E}\kern-.125emX}}

\begin{document}

\title{
EntroGD: Scalable Generalized Deduplication for Efficient Direct Analytics on Compressed IoT Data
\thanks{This work was supported in part by the GROWLean Project under Grant DFF-2035-00229B granted by the Independent Research Fund Denmark and by a Villum Synergy Research Grant (VIL 50075) from Villum Fonden.}
}


\author{
\IEEEauthorblockN{Xiaobo Zhao, Daniel E. Lucani}
\IEEEauthorblockA{
DIGIT, Department of Electrical and Computer Engineering,
Aarhus University\\
\{xiaobo.zhao, daniel.lucani\}@ece.au.dk}
}

\maketitle

\begin{abstract}
Massive data streams from IoT and cyber-physical systems must be processed under strict bandwidth, latency, and resource constraints.
Generalized Deduplication (GD) is a promising lossless compression framework, as it supports random access and direct analytics on compressed data.
However, existing GD algorithms exhibit quadratic complexity $\mathcal{O}(nd^{2})$, which limits their scalability for high-dimensional datasets.
This paper proposes \textbf{EntroGD}, an entropy-guided GD framework that decouples analytical fidelity from compression efficiency to achieve linear complexity $\mathcal{O}(nd)$.
EntroGD adopts a two-stage design, first constructing compact condensed samples to preserve information critical for analytics, and then applying entropy-based bit selection to maximize compression.
Experiments on 18 IoT datasets show that EntroGD reduces configuration time by up to $53.5\times$ compared to state-of-the-art GD compressors.
Moreover, by enabling analytics with access to only $2.6\%$ of the original data volume, EntroGD accelerates clustering by up to $31.6\times$ with negligible loss in accuracy.
Overall, EntroGD provides a scalable and system-efficient solution for direct analytics on compressed IoT data.
\end{abstract}

\section{Introduction}
\label{sec:introduction}

The explosive growth of data generated by Internet of Things (IoT) devices, cyber-physical systems, and edge platforms poses increasing challenges for efficient storage, transmission, and analytics~\cite{abdalzaher2025quality,jiang2018data,xu2023edge}.
Lossless compression methods such as Bzip2~\cite{bzip2}, LZ4~\cite{collet_lz4}, Snappy~\cite{ghemawat_snappy}, zlib~\cite{gailly_zlib}, and Zstd~\cite{collet_zstd} are widely used to reduce data volume.
However, these approaches typically require full decompression before analysis, which introduces additional latency and computational overhead, and limits their applicability to real-time or resource-constrained analytics workloads at the network edge.

To address these limitations, generalized deduplication (GD) has emerged as a promising compression framework that not only reduces storage and transmission costs, but also supports random access and direct analytics on compressed data~\cite{vestergaard2019_a,Vestergaard_2019b,Vestergaard_2020,vestergaard2021titchy}.
GD extends traditional deduplication by grouping similar, rather than identical, data chunks into bases and deviations, enabling lossless reconstruction while allowing approximate analytics to operate directly on compact base representations.
As a result, only a small fraction of the compressed stream, e.g., less than 2\%, needs to be accessed for analytics~\cite{hurst2024greedygd}.
This property makes GD particularly attractive for networked IoT and edge analytics pipelines, where minimizing data movement and processing delay is critical.
Recent work has further demonstrated the relevance of GD in networking and systems contexts, including stream aggregation~\cite{Aoshima2025}, privacy-preserving fog computing~\cite{Zhang2022}, and hybrid ARQ mechanisms for IoT communications~\cite{Xu2022}.

Despite these advantages, the performance of GD-based methods depends critically on how base bits are selected, since this choice determines both the number of unique bases and the analytical fidelity of the compressed representation.
Selecting optimal base bits is computationally challenging due to the exponential search space, for example $2^{32d}$ for 32-bit, $d$-dimensional data, which becomes prohibitive for high-dimensional datasets commonly encountered in modern sensing and monitoring applications.
Early approaches relied on inter-bit correlation analysis to select informative base bits~\cite{Vestergaard_2020}.
More recently, GreedyGD~\cite{hurst2024greedygd} introduced a greedy bit-selection strategy that iteratively minimizes the number of newly created bases during compression, using a tree-based structure (BASETREE) to balance compression efficiency and analytical accuracy.
GreedyGD has been refined for floating-point data~\cite{Taurone_2023}, extended to image compression~\cite{Rask_2024}, and shown strong performance in edge analytics tasks such as clustering~\cite{hurst2024greedygd} and anomaly detection~\cite{Taurone_2024}.

However, GD-based algorithms still face two fundamental challenges.
First, the iterative bit-selection process in GreedyGD incurs quadratic complexity $\mathcal{O}(nd^2)$, where $n$ is the number of samples and $d$ is the data dimensionality, which limits scalability for large-scale and high-dimensional IoT datasets.
Second, GreedyGD tightly couples two competing objectives, maximizing compression efficiency and preserving analytical fidelity.
Allocating more information to bases improves analytical accuracy but degrades compression, whereas prioritizing compression can significantly harm downstream analytics.
This coupling makes it difficult to scale GD to networked edge scenarios that require both efficiency and reliable analytics.

\begin{figure}[t]
    \centering
    \includegraphics[width=0.85\linewidth]{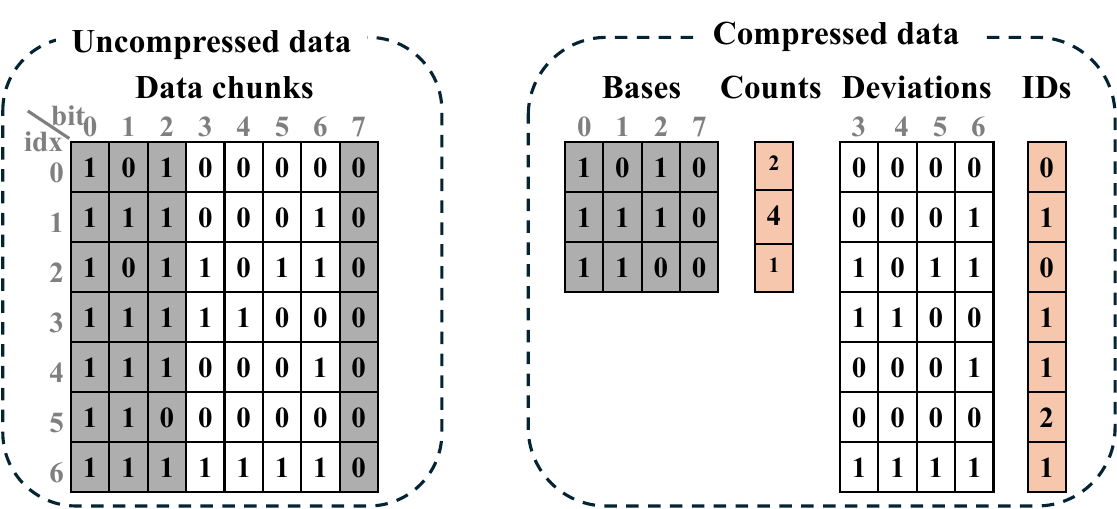}
    \caption{An example of Generalized Deduplication applied to 8-bit data chunks.
    The first three and the last bit are deduplicated as the base, while the remaining bits are stored verbatim as deviations with an ID linking them to the base.
    \vspace{-0.1in}
    }
    \label{fig:gd}
\end{figure}

To overcome these limitations, we propose EntroGD, an entropy-guided GD-based compression framework that decouples analytics and compression into two distinct stages.
In the first stage, EntroGD identifies base bits that preserve information critical for analytics and constructs a small set of condensed samples that summarize intra-base variations, enabling accurate analytics on a compact representation.
In the second stage, EntroGD applies entropy-guided bit selection to maximize compression efficiency, eliminating the need for iterative search and reducing computational complexity from $\mathcal{O}(nd^2)$ to $\mathcal{O}(nd)$.


Our main contributions are summarized as follows:
\begin{itemize}
    \item We identify scalability and objective-coupling limitations in existing GD algorithms, and address these limitations by proposing EntroGD, an entropy-guided GD framework that decouples analytics fidelity from compression optimization into two distinct stages.
    
    \item We introduce a condensed sample generation mechanism that preserves critical information via compact, weighted representations, improving analytical accuracy on compressed data compared to existing GD-based approaches.

    \item We design a non-iterative, entropy-guided base bit selection strategy that reduces the computational complexity of GD configuration from $\mathcal{O}(nd^2)$ to $\mathcal{O}(nd)$.
    
    \item We conduct extensive evaluations on 18 diverse IoT datasets, demonstrating that EntroGD achieves compression ratios comparable to state-of-the-art compressors, reduces configuration time by up to $53.5\times$, and accelerates clustering by up to $31.6\times$ while accessing only $2.6\%$ of the original data volume, enabling low-latency processing and scalable deployment on resource-constrained devices handling large, high-dimensional datasets.
\end{itemize}

\section{Background}
\label{sec:background}

\subsection{Generalized Deduplication}
\label{sec:generalized_deduplication}

As illustrated in Fig.~\ref{fig:gd}, 
GD splits data chunks into subsets of bits with frequently appearing bit patterns, \emph{bases}, and subsets of bits with high-variances, \emph{deviations}. GD deduplicates bases and stores deviations unchanged alongside pointers to corresponding bases to enable lossless decompression. 
The base table and base counts serve as a compact data summary, enabling approximate analytics directly on them.
Effective compression occurs when the number of unique bases is small relative to the total number of data chunks.

\subsection{GreedyGD Algorithm}
\label{sec:greedygd}

GreedyGD is a state-of-the-art GD-based compression algorithm~\cite{hurst2024greedygd}.
It first scales floating-point data and converts them to integers, a preprocessing step that improves compressibility.
Using the \textsc{BaseTree} algorithm, GreedyGD tracks the number of bases during bit selection in $\mathcal{O}(n)$ time.
At each iteration, the algorithm traverses all $d$ dimensions, calculates the number of bases, evaluates the cost of promoting the most significant non-base bit in each dimension, and selects the bit with the lowest cost as the next base bit.
The cost function jointly considers compression ratio and analytical performance.
This process continues until no further cost reduction is achieved, resulting in an overall time complexity of $\mathcal{O}(nd^2)$.

When performing analytics directly on compressed data, only the bases and their counts are accessed, avoiding full decompression.
To reduce approximation error, the base centroid $b_c$ is used, defined as
\begin{equation}
    b_c = (b + b_{\max})/2,
\label{eq:base_centroid}
\end{equation}
where $b$ denotes the value of a base and $b_{\max}$ represents its maximum attainable value.
These are obtained by filling the deviation bits with all zeros and all ones, respectively, converting the resulting binary representations to decimal, and, if necessary, casting from integer to floating-point.
Note that $b_c$, $b$, and $b_{\max}$ are $d$-dimensional vectors for multidimensional data, where $d > 1$.
The base counts are used as weights of the centroids when performing analytics.

\section{EntroGD Algorithm}
\label{sec:entroGD}

\begin{figure}[!t]
    \centering
    \footnotesize
    \renewcommand{\arraystretch}{1.1}

\begin{tabular}{p{\algowidth}}
	\toprule
	\showAlgoCounter{\label{alg:condensed_samples_alg}}{Condensed Sample Generation in EntroGD} \\ \midrule
	\textbf{Inputs:} dataset $ \mathcal{D} $, threshold of the number of condensed samples $m_{\max}$ \\
	\textbf{Outputs:} $m$ condensed samples with weights $\{s_j, w_j\}$, $j=1, 2, \ldots, m$ \\
	\vspace{-0.65em}
	\begin{algorithmic}[1]
		\State $ \basebits \gets $ constant bits; $ m \gets 1$ if $ \basebits \neq \emptyset $ else 0
		\While{\(m < m_{\max}\) and not all bits in $ \basebits $}
            \State \(C \gets\) one left-most remaining bit per dimension
            \For{each bit \(i \in C\)}
                \State \(B \gets B \cup \{i\}\); $m \gets $ \textsc{BaseTree} count
                \If{\(m \ge m_{\max}\)} \textbf{break} \EndIf
            \EndFor
        \EndWhile

        \State obtain bases and chunk indices \(\{b_j, I_j\}_{j=1}^{m}\) using \textsc{BaseTree}
        \For{\(j = 1 \to m\)}
            \State \(b_j \gets \) base value; \(w_j \gets |I_j|\); \(\mu_j \gets \frac{1}{w_j}\sum_{k \in I_j} \delta_{j,k}\)
            \State \(s_j \gets b_j + \mu_j\)
        \EndFor

		\State \Return $\{s_j, w_j\}_{j=1}^{m}$
	\end{algorithmic}
	\\[-1em] \bottomrule
\end{tabular}

\end{figure}

\begin{figure*}[t]
    \centering
    \includegraphics[width=\textwidth]{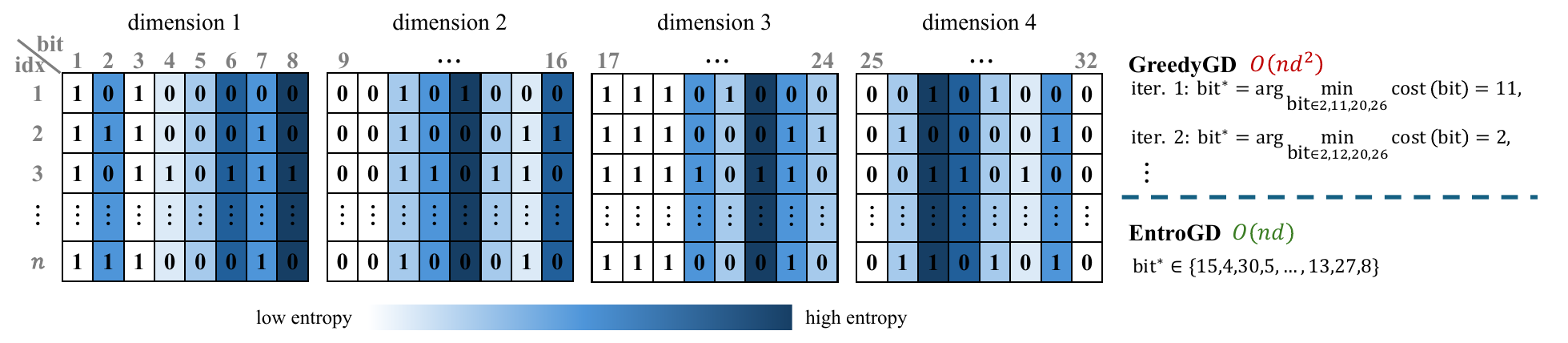}
    \caption{Comparison of base bit selection in GreedyGD and EntroGD on a dataset with $n$ samples and $d = 4$ dimensions of 8-bit data.
    GreedyGD iteratively selects $\text{bit}^*$ that minimize the cost function as base bits, whereas EntroGD selects them in ascending order of entropy.
    By eliminating the iterative search, EntroGD reduces the complexity from $\mathcal{O}(nd^2)$ to $\mathcal{O}(nd)$.\vspace{-0.1in}}
    \label{fig:greedy_vs_entro}
\end{figure*}

This section introduces EntroGD, which separates the competing objectives of analytics and compression in GD-based algorithms into two stages, avoiding mutual compromise.
EntroGD first selects base bits that preserve information critical for analytics and generates condensed samples from the bases and their deviations for analytical use.
These samples are appended to the original data to form an extended dataset, which is then compressed with the sole objective of maximizing compression.
Since the condensed samples and their weights are small relative to the original data, the overall compression is only marginally affected despite the dataset expansion.

\subsection{Preliminaries}
\label{sec:preliminaries}

We consider a dataset with $n$ samples, each of dimension $d$.
If the data are floating-point, they are preprocessed as in GreedyGD.
For each sample, concatenating the per-dimension binary representations yields a binary chunk of length $l_c$ bits.
Then, we calculate the entropy of each bit position across all $n$ chunks.
The entropy of non-constant bit position $i$ is
\begin{equation}
    H(i) = -p_i \log_2(p_i) - (1 - p_i) \log_2(1 - p_i),
\label{eq:entropy}
\end{equation}
where $p_i$ denotes the (estimated) probability that bit $i$ equals 1 across all data chunks, and $1 \leq i \leq l_c$.
The entropy of a constant bit position is 0.

\subsection{Condensed Sample Generation}
\label{sec:condensed_samples}

As outlined in Algorithm~\ref{alg:condensed_samples_alg}, EntroGD first generates $m$ condensed samples for analysis with a maximum threshold $m_{\max}$.
To retain critical information and maintain balance across feature dimensions, the leftmost non-constant $d$ bits, one for each dimension, are first considered.
If all features are equally important, bits are selected sequentially, otherwise, they are chosen by feature importance in descending order.
The process then repeats for the next set of remaining leftmost $d$ bits.
After each selection, we use the \textsc{BaseTree} algorithm~\cite{hurst2024greedygd} to count the number of unique bases (condensed samples).
The selection stops once $m$ exceeds $m_{\max}$.

This procedure ensures that the selected bits capture informative content across all dimensions, which is essential for maintaining analytical accuracy.
Furthermore, it provides explicit control over the number of bases, and consequently the size of the condensed samples used for analytics, a feature not supported by GreedyGD or other GD-based algorithms.

With $m$ bases identified, and their corresponding deviations, we generate $m$ condensed samples for analytics.
For each base, the mean of its deviations is computed and added to the base to form a condensed sample $s_j$, as follows
\begin{equation}
    s_j = b_j + \frac{1}{w_j} \sum_{k \in I_j} \delta_{j,k},
\label{eq:condensed_sample}
\end{equation}
where $b_j$ denotes the value of the $j$-th base, $\delta_{j,k}$ is its $k$-th deviation, $I_j$ is the corresponding deviation index set, and $w_j = |I_j|$ is the number of associated deviations, for $1 \leq j \leq m$.
$b_j$ is obtained by setting all deviation bits to zero and converting the result to decimal, with an optional cast to floating-point.  
Similarly, \({\delta}_{j,k}\) is computed by zeroing the base bits of a deviation and performing the same conversion.
The values of $s_j$, $b_j$, and \({\delta}_{j,k}\) are vectors of dimension \(d\) for multidimensional data, where \(d > 1\).

The generated samples effectively capture the average characteristics of the data chunks associated with each base, and reduce analytical error compared with using the base centroid in GreedyGD, as the mean deviation offers a more accurate representation of the data distribution within each base.
Each sample is assigned a weight $w_j$, thereby reflecting its relative importance in subsequent analysis.
This is equivalent to the \emph{counts} in Fig.~\ref{fig:gd}.
These $m$ samples are appended to the original dataset to form an extended set of $n + m$ samples for further compression, while weights are stored separately.

\subsection{Compression}
\label{sec:compression}

As mentioned previously, the key challenge in GD-based algorithms has focused on determining how to split data chunks into bases and deviations to minimize the number of unique bases while retaining sufficient information in bases for analytical fidelity.

In contrast, EntroGD uses the condensed samples for analytics, allowing the compression stage to focus solely on maximizing compression.
As illustrated in Fig.~\ref{fig:greedy_vs_entro}, EntroGD selects base bits using entropy guidance: all bit positions are sorted in ascending order of entropy, and low-entropy bits are chosen first.
For simplicity, entropy values are computed once from the original data and not recalculated after appending condensed samples.  
Although not guaranteed to be optimal, this approach typically results in a small number of unique bases.
Constant bits, an extreme case with zero entropy, always promote base reuse when included in the base bit set.
Theoretical analysis of this mechanism is left for future work.

During compression, after each bit selection, the \textsc{BaseTree} algorithm counts the number of bases $n_b$, and the compressed size $S$ is computed as
\begin{equation}
S = n_b l_b + (n + m)(l_d + l_{id}) + m l_w + S_{\text{params}},
\label{eq:compressed_size}
\end{equation}
where $l_b$ and $l_d$ are the bit lengths of the base and deviation, respectively, with $l_b + l_d = l_c$,
$l_w = \lceil \log_2 n \rceil$ is the bit length of the condensed sample weight,
$l_{id} = \lceil \log_2 n_b \rceil$ is the bit length of the base index, and
$S_{\text{params}}$ denotes the size of the auxiliary parameters, typically negligible.
The selection process ends when the compressed size does not decrease for $\tau$ consecutive selections or when all bits have been processed.
In our experiments, the plateau threshold is set to $\tau = 10$.
The compression process is summarized in Algorithm~\ref{alg:compression_alg}.

\begin{figure}[!t]
	\centering
	\footnotesize
	\renewcommand{\arraystretch}{1.1}

\begin{tabular}{p{\algowidth}}
	\toprule
	\showAlgoCounter{\label{alg:compression_alg}}{Compression in EntroGD} \\ \midrule
	\textbf{Inputs:} extended dataset $ \mathcal{D}' $, entropy values $H$, plateau threshold $\tau$ \\
	\textbf{Outputs:} base bits $ \basebits $, bases $\mathbf{B}$, deviations $\boldsymbol{\Delta}$ with base IDs $\mathbf{ID}$\\
	\vspace{-0.65em}
	\begin{algorithmic}[1]
		\State $ \basebits_{\text{best}} \gets $ constant bits; $ \basebits \gets \basebits_{\text{best}} $; $ L \gets $ bits sorted by $H$
        \State $ S_{\text{best}} \gets $ compressed size (Eq.~\eqref{eq:compressed_size}); $ \tau_{\text{count}} \gets 0 $
		\For{each bit \(i \in L\)}
            \State \(B \gets B \cup \{i\}\); $n_b \gets $ \textsc{BaseTree}; $S \gets$ size (Eq.~\eqref{eq:compressed_size})
            \If{\(S < S_{\text{best}}\)} $S_{\text{best}} \gets S$; $\basebits_{\text{best}} \gets \basebits$; $\tau_{\text{count}} \gets 0$
            \Else \ $\tau_{\text{count}} \gets \tau_{\text{count}} + 1$
            \EndIf
            \State \textbf{if} \(\tau_{\text{count}} \ge \tau\) \textbf{then break}
        \EndFor
        \State \Return $ \basebits_{\text{best}} $ and GD result ($\mathbf{B}$, $\boldsymbol{\Delta}, \mathbf{ID}$) on $ \mathcal{D}' $
	\end{algorithmic}
	\\[-1em] \bottomrule
\end{tabular}

\end{figure}

\subsection{Complexity Analysis}
\label{sec:complexity_analysis}



EntroGD employs a non-iterative, entropy-guided strategy for base bit selection, reducing the overall computational complexity to $\mathcal{O}(nd)$.
To detail the complexity, we consider each major step of the algorithm.
In the entropy computation phase, calculating entropy across all $l_c$ bit positions for $n$ samples requires $\mathcal{O}(nl_c)$.
During condensed sample generation, producing up to $m_{\max}$ samples involves examining at most $l_c$ bits and performing base counting with \textsc{BaseTree} for each, costing $\mathcal{O}(n)$ per bit and yielding a total complexity of $\mathcal{O}(nl_c)$.
Computing mean deviations and forming condensed samples adds another $\mathcal{O}(nd)$.
In the compression stage, up to $l_c$ bits are added to $B$, each requiring base counting via \textsc{BaseTree} ($\mathcal{O}(n)$ per bit), again resulting in $\mathcal{O}(nl_c)$.
With the plateau threshold $\tau$, the number of bits added is usually much smaller than $l_c$, giving a practical cost below $\mathcal{O}(nl_c)$.
Once the base bits are identified, applying GD to the extended dataset to derive $\mathbf{B}$, $\boldsymbol{\Delta}$, and $\mathbf{ID}$ requires scanning the entire binary dataset, also incurring $\mathcal{O}(nl_c)$.
Therefore, the total computational complexity of EntroGD is $\mathcal{O}(4nl_c + nd)$, which simplifies to $\mathcal{O}(nl_c)$.
Since $l_c$ is proportional to $d$ (e.g., $l_c = 32d$ for 32-bit data), the overall complexity can be expressed as $\mathcal{O}(nd)$.

\section{Performance Evaluation}
\label{sec:evaluation}

\subsection{Experimental Setup}
\label{sec:setup}

We evaluate EntroGD on 18 datasets (Table~\ref{tab:clust_datasets}) covering diverse sizes, dimensionalities, data types, and precisions.
The comparison includes GreedyGD, GreedyGD+, EntroGD, and universal compressors (Bzip2, LZ4, Snappy, zlib, and Zstd) configured to their maximum compression level.
GreedyGD follows the default configuration in~\cite{hurst2024greedygd}.
GreedyGD+ extends GreedyGD by storing additional deviation means for the corresponding bases as we propose for EntroGD in Section~\ref{sec:condensed_samples}.
In EntroGD, $m_{\max}$ is set to the number of bases in GreedyGD, and the condensed samples are truncated accordingly for a fair analytical comparison. 
We perform $k$-means data clustering as our analytics task and carry it out on the original data and three GD compressed representations.
All experiments are conducted on a MacBook Pro with an Apple M3 Pro chip, 18~GB memory, Python~3.12.1, and scikit-learn~1.7.2.

\begin{table}[t]
    \centering
    \caption{Datasets Used in Experiments.}
    \label{tab:clust_datasets}
    \resizebox{\columnwidth}{!}{%
    \tiny
    \
    \begingroup
    \setlength{\tabcolsep}{4pt}
    \renewcommand{\arraystretch}{1.15}
    \begin{tabular}{llrrrr}
        \toprule
        \textbf{Dataset} & \textbf{Type} & \textbf{Precision} & $\boldsymbol{n}$ & $\boldsymbol{d}$ & \textbf{Size} (kB) \\
        \midrule
        Aarhus Citylab~\cite{AarhusKommune_2017}        & float & 32-bit & 26\,387      & 4  & 422      \\
        Aarhus Pollution 172156~\cite{Ali_2015}         & int   & 32-bit & 17\,568      & 5  & 351      \\
        Aarhus Pollution 204273~\cite{Ali_2015}         & int   & 32-bit & 17\,568      & 5  & 351      \\
        Chicago Beach Water I~\cite{CityChicago_2022b}  & float & 32-bit & 39\,829      & 5  & 797      \\
        Chicago Beach Water II~\cite{CityChicago_2022b} & float & 32-bit & 10\,034      & 6  & 241      \\
        Chicago Beach Weather~\cite{CityChicago_2022a}  & float & 32-bit & 86\,694      & 9  & 3\,121    \\
        Chicago Beach Weather~\cite{CityChicago_2022a}  & int   & 32-bit & 86\,763      & 5  & 1\,735    \\
        Chicago Taxi Trips~\cite{CityChicago_2022c}     & float & 64-bit & 3\,466\,498  & 10 & 277\,320  \\
        CMU IMU acceleration~\cite{Torre_2009}          & float & 32-bit & 134\,435     & 3  & 1\,613    \\
        CMU IMU Velocity~\cite{Torre_2009}              & float & 32-bit & 134\,435     & 3  & 1\,613    \\
        CMU IMU Magnetic~\cite{Torre_2009}              & float & 32-bit & 134\,435     & 3  & 1\,613    \\
        CMU IMU Position~\cite{Torre_2009}              & float & 32-bit & 134\,435     & 4  & 2\,151    \\
        CMU IMU All~\cite{Torre_2009}                   & float & 32-bit & 134\,435     & 13 & 6\,991    \\
        COMBED Mains Power~\cite{Batra_2014}            & float & 64-bit & 82\,888      & 3  & 995      \\
        COMBED UPS Power~\cite{Batra_2014}              & float & 64-bit & 86\,199      & 3  & 1\,035    \\
        Melbourne City Climate~\cite{CityMelbourne_2020}& float & 32-bit & 56\,570      & 3  & 679      \\
        Gas Turbine Emissions~\cite{Kaya_2019}          & float & 32-bit & 36\,733      & 11 & 1\,616    \\
        Household Power Usage~\cite{Hebrail_2012}       & float & 32-bit & 2\,049\,280  & 7  & 57\,380   \\
        \bottomrule
    \end{tabular}
    \endgroup
    }%
\end{table}

\begin{figure}[t]
    \centering
    \includegraphics[width=0.8\linewidth]{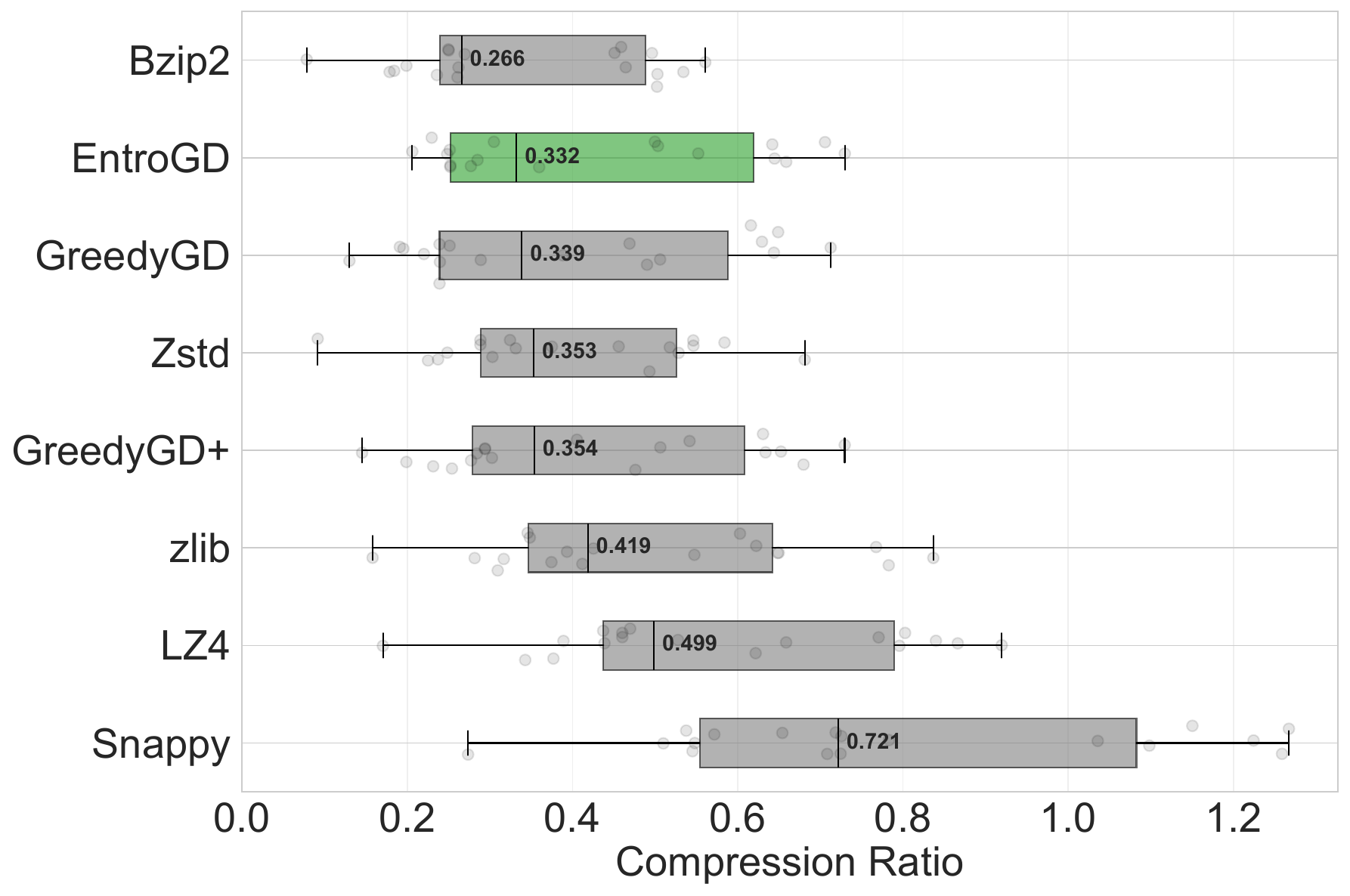}
    \caption{Box plot of CR across all datasets. EntroGD achieves the second-lowest median CR after Bzip2.
    \vspace{-0.2in}
    }
    \label{fig:compression_ratios_boxplot}
\end{figure}

\begin{figure*}[t]
    \centering
    \begin{subfigure}[b]{0.18\linewidth}
        \centering
        \includegraphics[width=\linewidth]{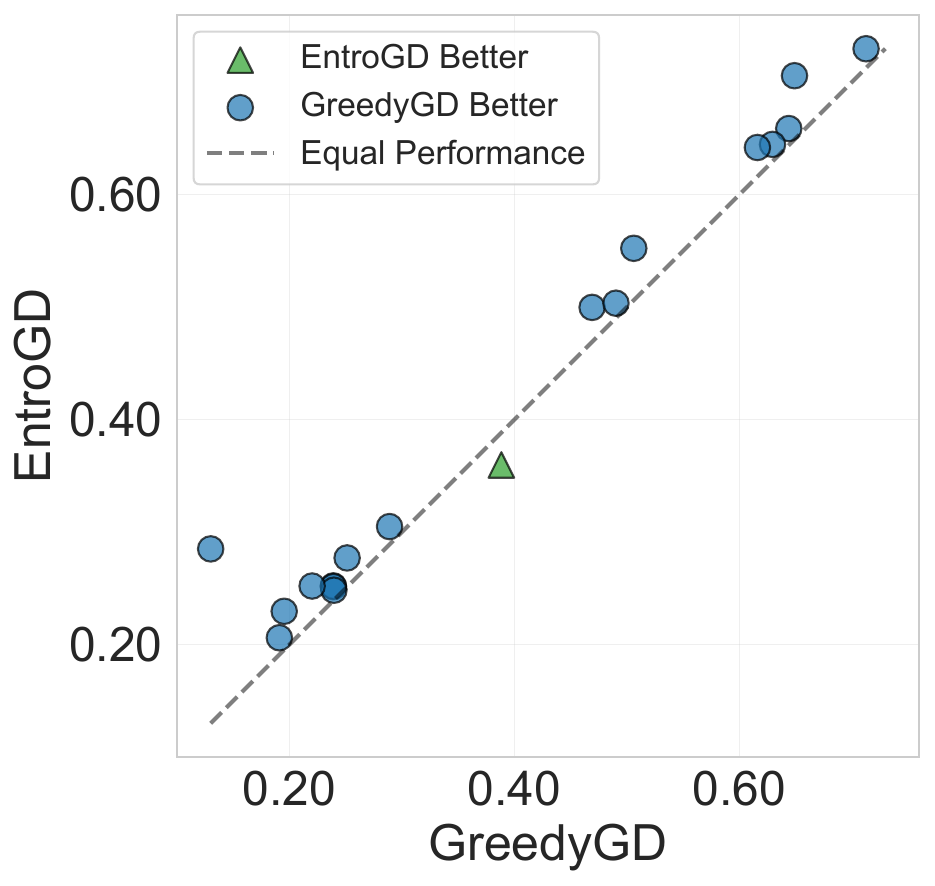}
        \caption{CR}
        \label{fig:compression_ratio_scatter}
    \end{subfigure}\hfill
    \begin{subfigure}[b]{0.18\linewidth}
        \centering
        \includegraphics[width=\linewidth]{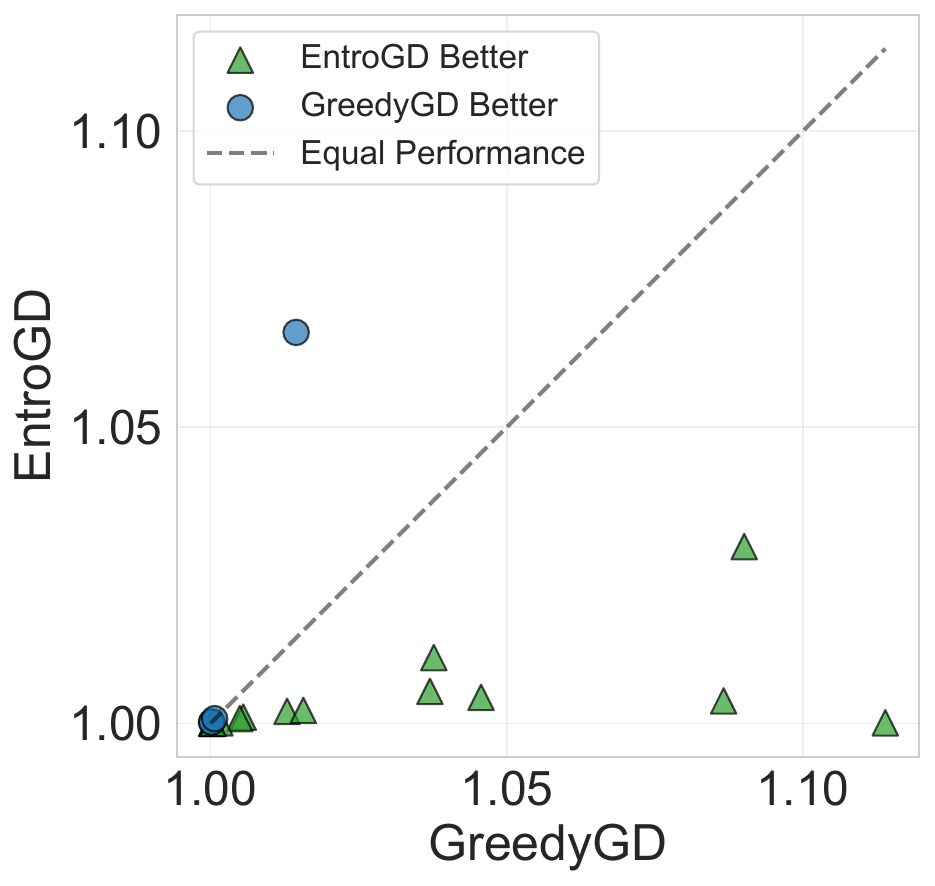}
        \caption{AR}
        \label{fig:approximation_ratio_scatter}
    \end{subfigure}\hfill
    \begin{subfigure}[b]{0.18\linewidth}
        \centering
        \includegraphics[width=\linewidth]{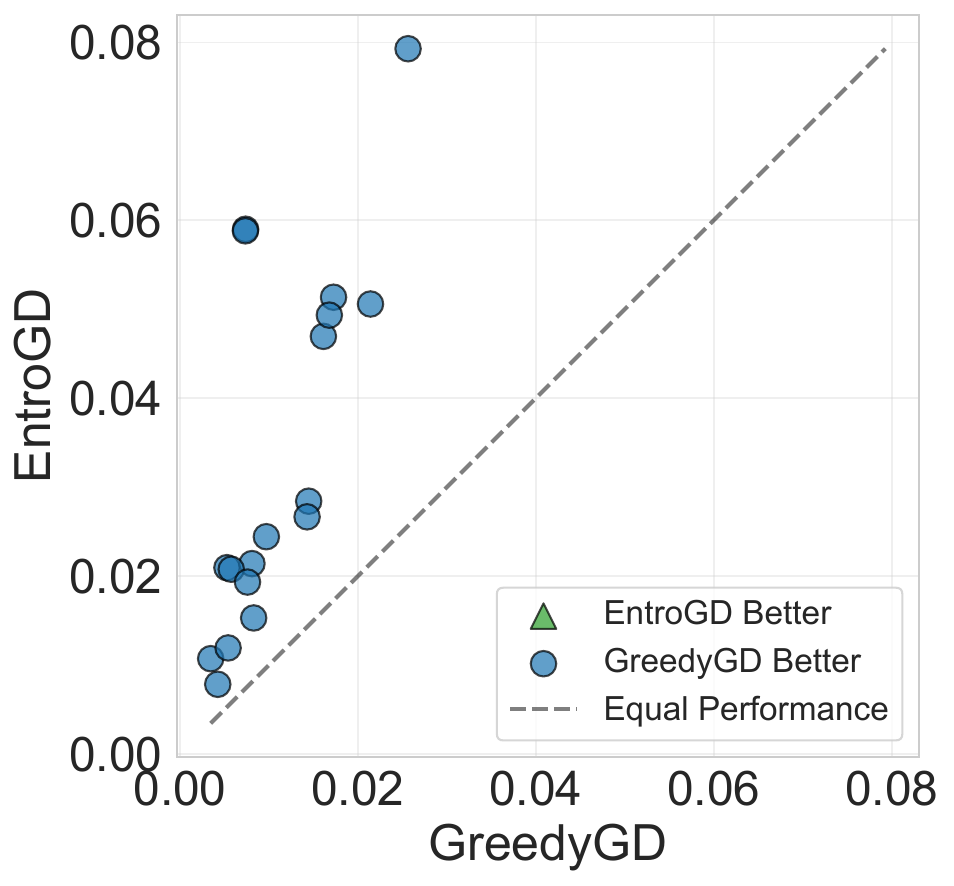}
        \caption{ADR}
        \label{fig:adr_scatter}
    \end{subfigure}\hfill
    \begin{subfigure}[b]{0.18\linewidth}
        \centering
        \includegraphics[width=\linewidth]{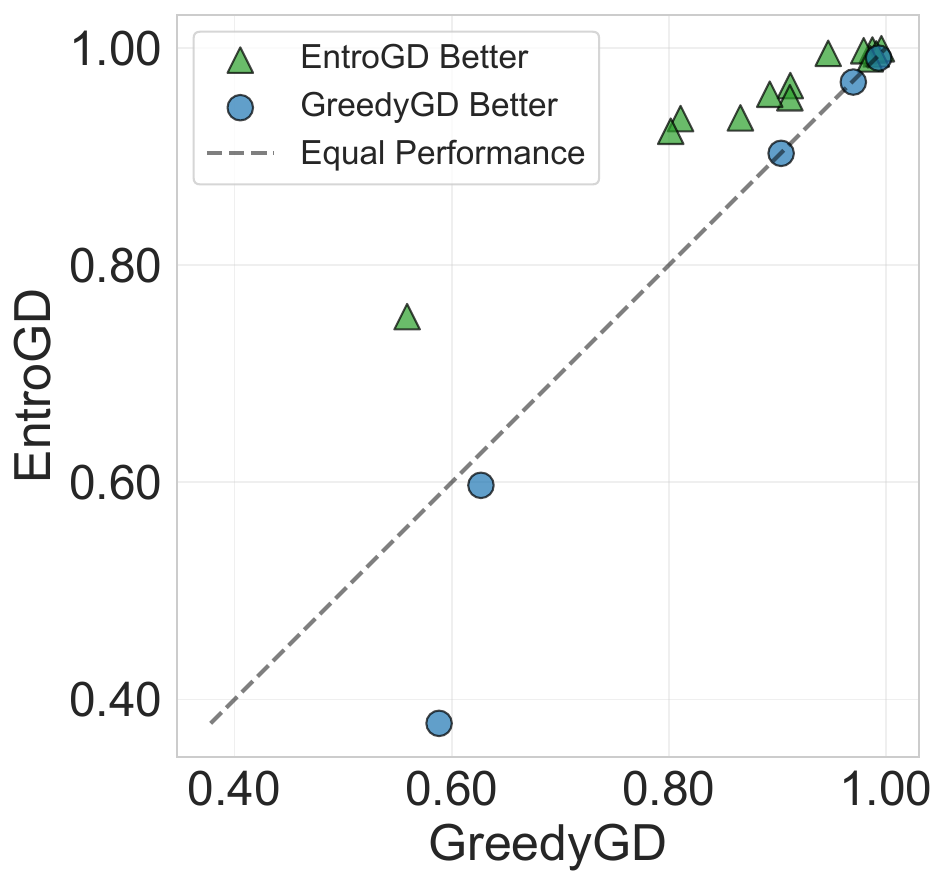}
        \caption{AMI}
        \label{fig:ami_scatter}
    \end{subfigure}\hfill
    \begin{subfigure}[b]{0.18\linewidth}
        \centering
        \includegraphics[width=\linewidth]{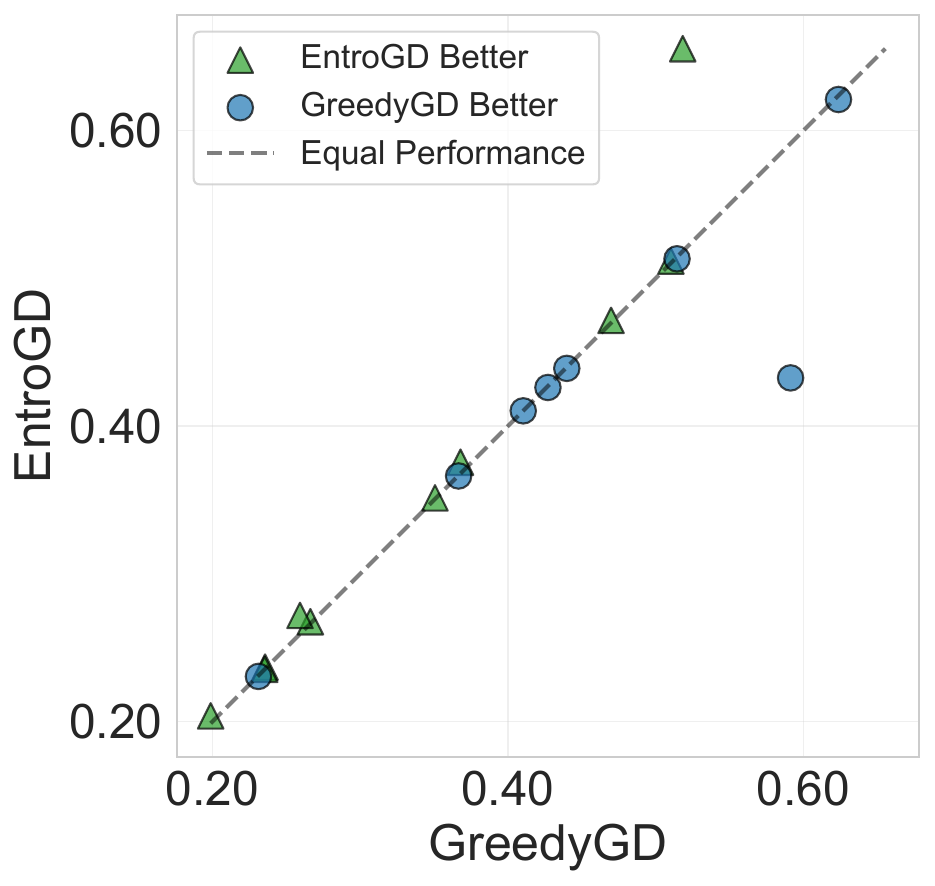}
        \caption{Silhouette}
        \label{fig:silhouette_scatter}
    \end{subfigure}
    \caption{Detailed performance comparison between EntroGD and GreedyGD across all datasets. 
    }
    \label{fig:metrics_scatter_side_by_side}
\end{figure*}

\begin{figure*}[t]
    \centering
    \begin{subfigure}[b]{0.18\linewidth}
        \centering
        \includegraphics[width=\linewidth]{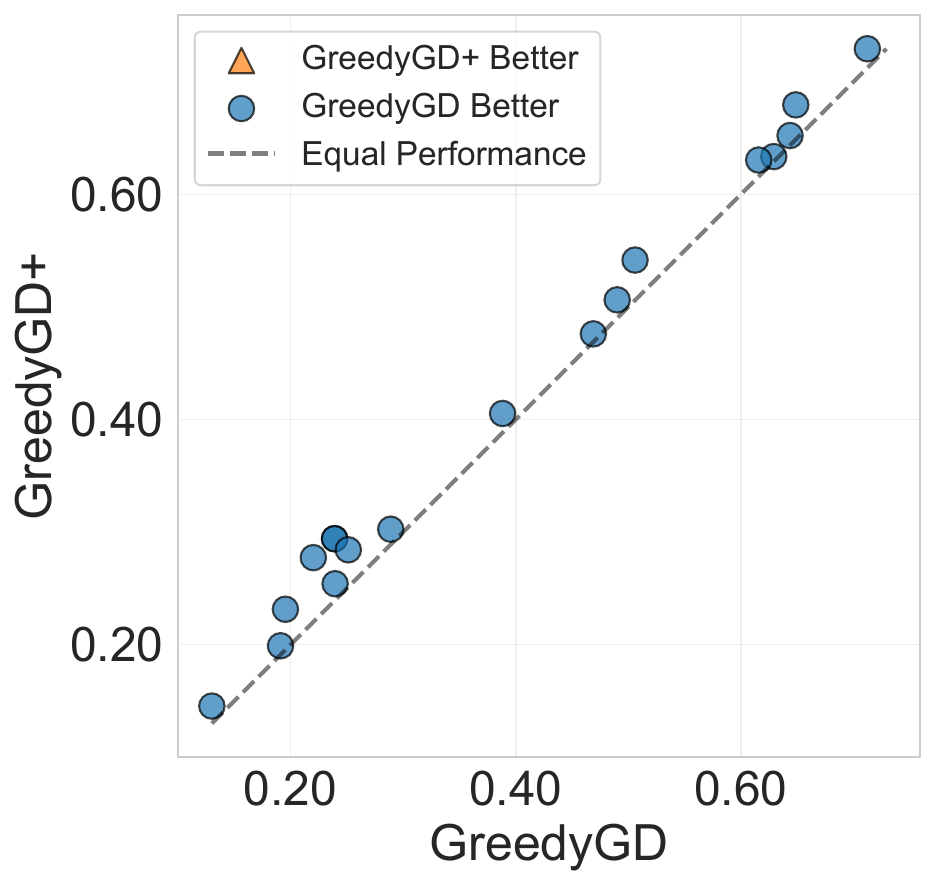}
        \caption{CR}
        \label{fig:compression_ratio_scatter_plus}
    \end{subfigure}\hfill
    \begin{subfigure}[b]{0.18\linewidth}
        \centering
        \includegraphics[width=\linewidth]{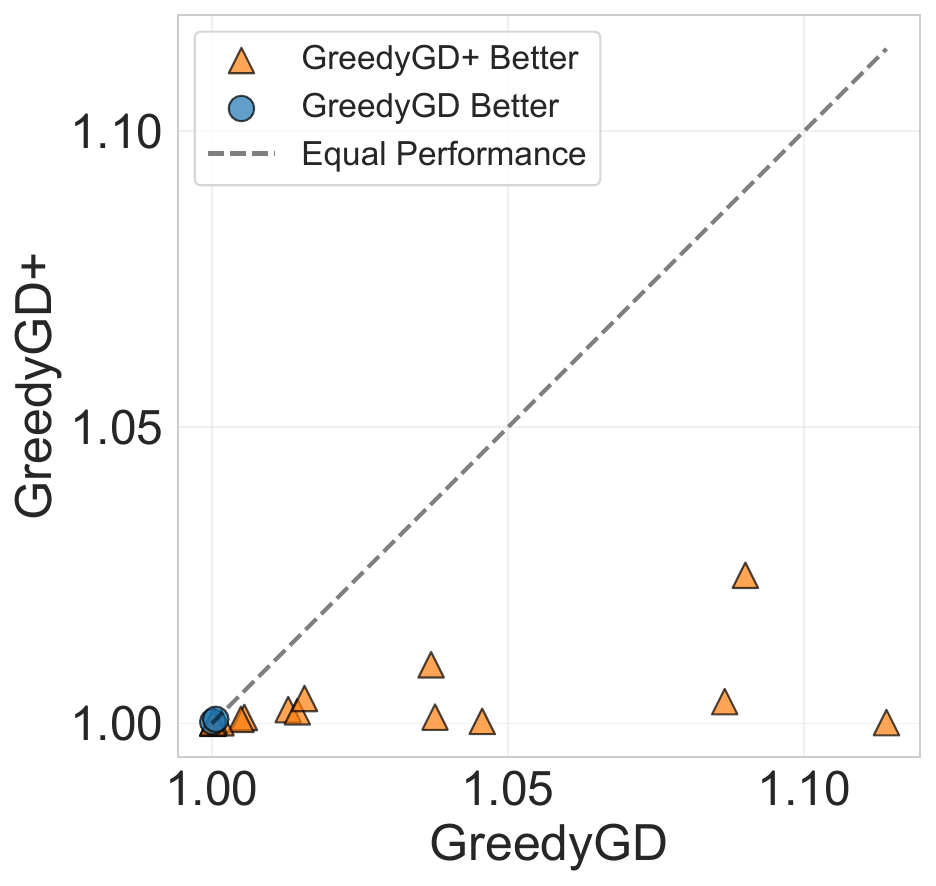}
        \caption{AR}
        \label{fig:approximation_ratio_scatter_plus}
    \end{subfigure}\hfill
    \begin{subfigure}[b]{0.18\linewidth}
        \centering
        \includegraphics[width=\linewidth]{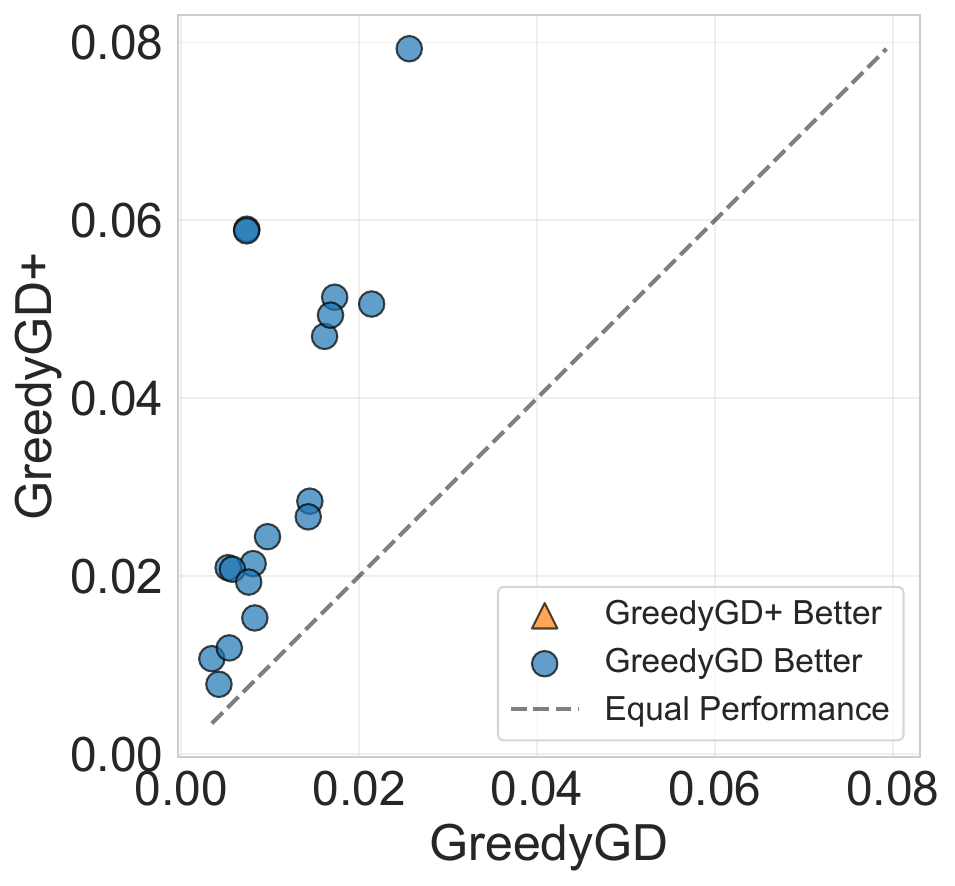}
        \caption{ADR}
        \label{fig:adr_scatter_plus}
    \end{subfigure}\hfill
    \begin{subfigure}[b]{0.18\linewidth}
        \centering
        \includegraphics[width=\linewidth]{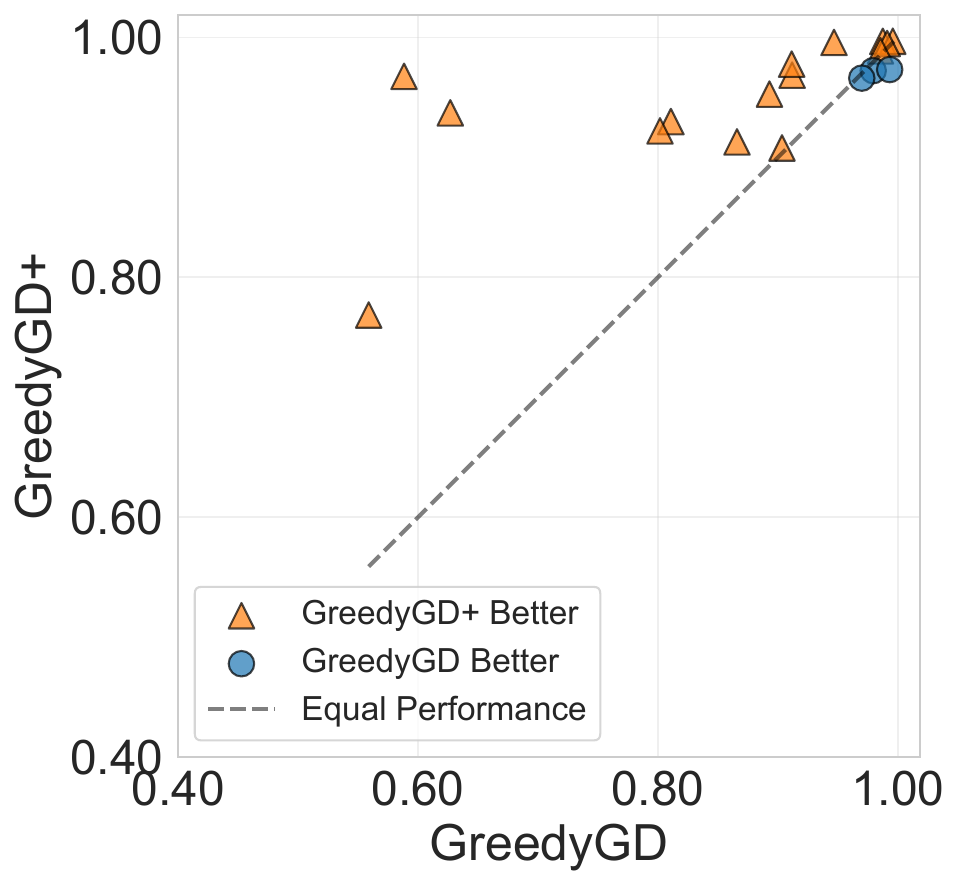}
        \caption{AMI}
        \label{fig:ami_scatter_plus}
    \end{subfigure}\hfill
    \begin{subfigure}[b]{0.18\linewidth}
        \centering
        \includegraphics[width=\linewidth]{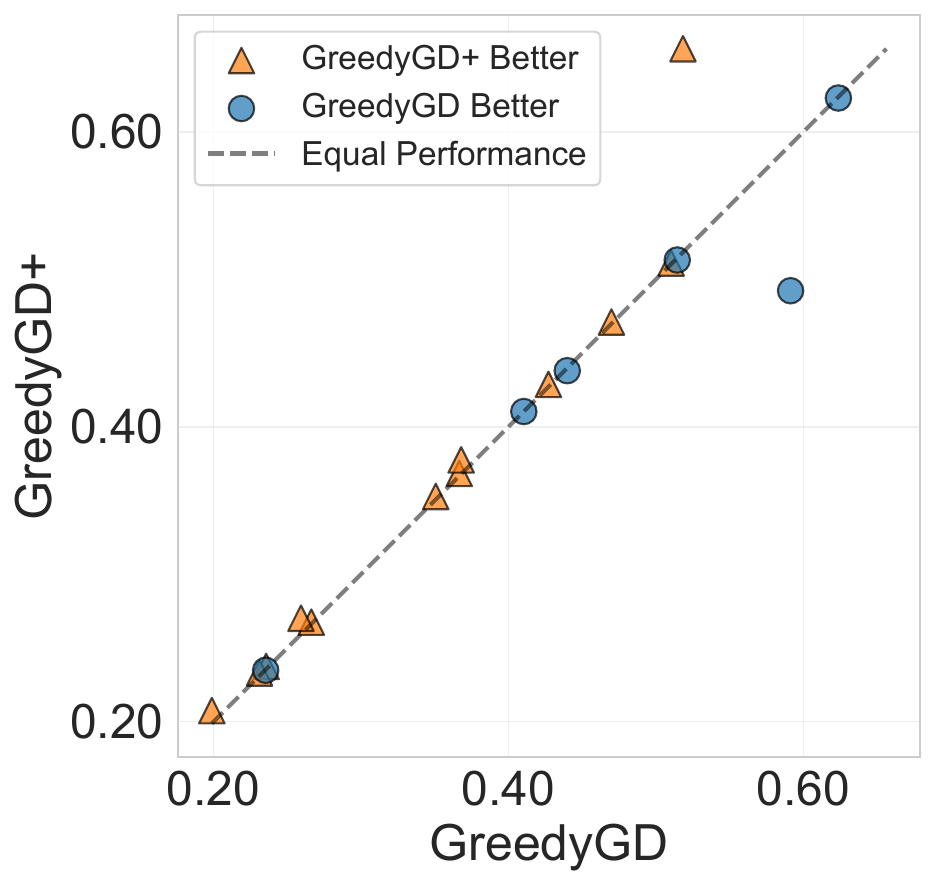}
        \caption{Silhouette}
        \label{fig:silhouette_scatter_plus}
    \end{subfigure}
    \caption{Detailed performance comparison between GreedyGD+ and GreedyGD across all datasets.
    }
    \label{fig:metrics_scatter_side_by_side_plus}
\end{figure*}
\subsection{Evaluation Metrics}
\label{sec:metrics}

Compression effectiveness is evaluated using the \emph{compression ratio} (CR), defined as the ratio of compressed to uncompressed data size, where lower values indicate better compression.
For GD-based methods, we also measure the \emph{configuration time}, i.e., the time required to determine the base bit set, which the most computationally intensive step.
We consider the following metrics to measure performance of $k$-means clustering on compressed data:
\begin{itemize}
\item \textbf{Approximation Ratio (AR)}: the ratio of the sum of squared errors in clustering on compressed data to the result obtained when clustering on the original data. Lower values are better and 1 is ideal.
\item \textbf{Analytics Data Ratio (ADR)}: the ratio of the size of compressed data accessed for analytics to the uncompressed data, where lower values are better.
\item \textbf{Adjusted Mutual Information (AMI)}: quantifies the similarity between clustering results on compressed and original data, where higher values are better and 1 is ideal.
\item \textbf{Silhouette Coefficient}: measures the cohesion and separation of clusters obtained from compressed data, where higher values indicate better clustering and 1 is ideal.
\item \textbf{Clustering Time}: time taken to perform clustering task.
\end{itemize}

\subsection{Compression Performance}
\label{sec:compression_performance}

\subsubsection{Compression Ratio}
\label{sec:compression_ratio}

Fig.~\ref{fig:compression_ratios_boxplot} shows the box plot of CR across all datasets.
Based on the median CR, EntroGD ranks second overall after Bzip2, demonstrating competitive compression efficiency.
Among GD-based methods, although GreedyGD achieves slightly lower CRs on most datasets (Fig.~\ref{fig:compression_ratio_scatter} and~\ref{fig:compression_ratio_scatter_plus}), EntroGD provides comparable compression with much higher efficiency. 

Compared to universal compressors, EntroGD outperforms zlib, LZ4 and Snappy, performs comparably to Zstd, and approaches Bzip2, which achieves the lowest median CR overall, but does not support random access or direct analytics.

\subsubsection{Configuration Time of GD-based Methods}
\label{sec:configuration_time}

\begin{figure}[t]
    \centering
    \includegraphics[width=0.8\linewidth]{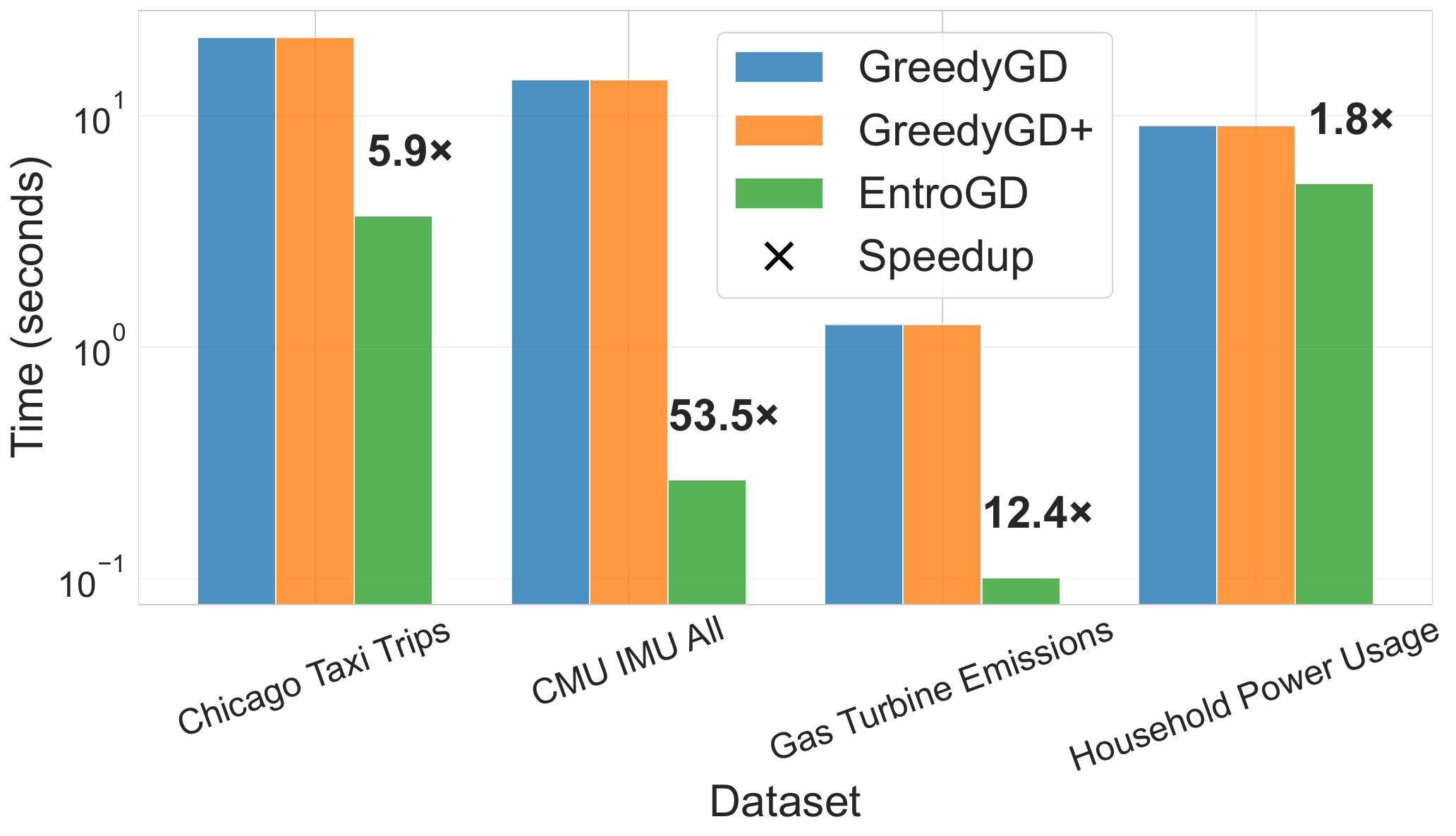}
    \caption{Configuration time comparison.
    \vspace{-0.1in}
    }
    \label{fig:compression_times_gd_methods}
\end{figure}

We chose four large and/or high-dimensional datasets (Chicago Taxi Trips, CMU IMU All, Gas Turbine Emissions, and Household Power Usage) to compare the configuration time of GD-based methods.
As shown in Fig.~\ref{fig:compression_times_gd_methods}, EntroGD achieves up to 53.5$\times$ speedup over GreedyGD/GreedyGD+, owing to its linear complexity $\mathcal{O}(nd)$ compared to the quadratic $\mathcal{O}(nd^2)$ of GreedyGD/GreedyGD+.

\begin{figure}[t]
    \centering
    \includegraphics[width=0.8\linewidth]{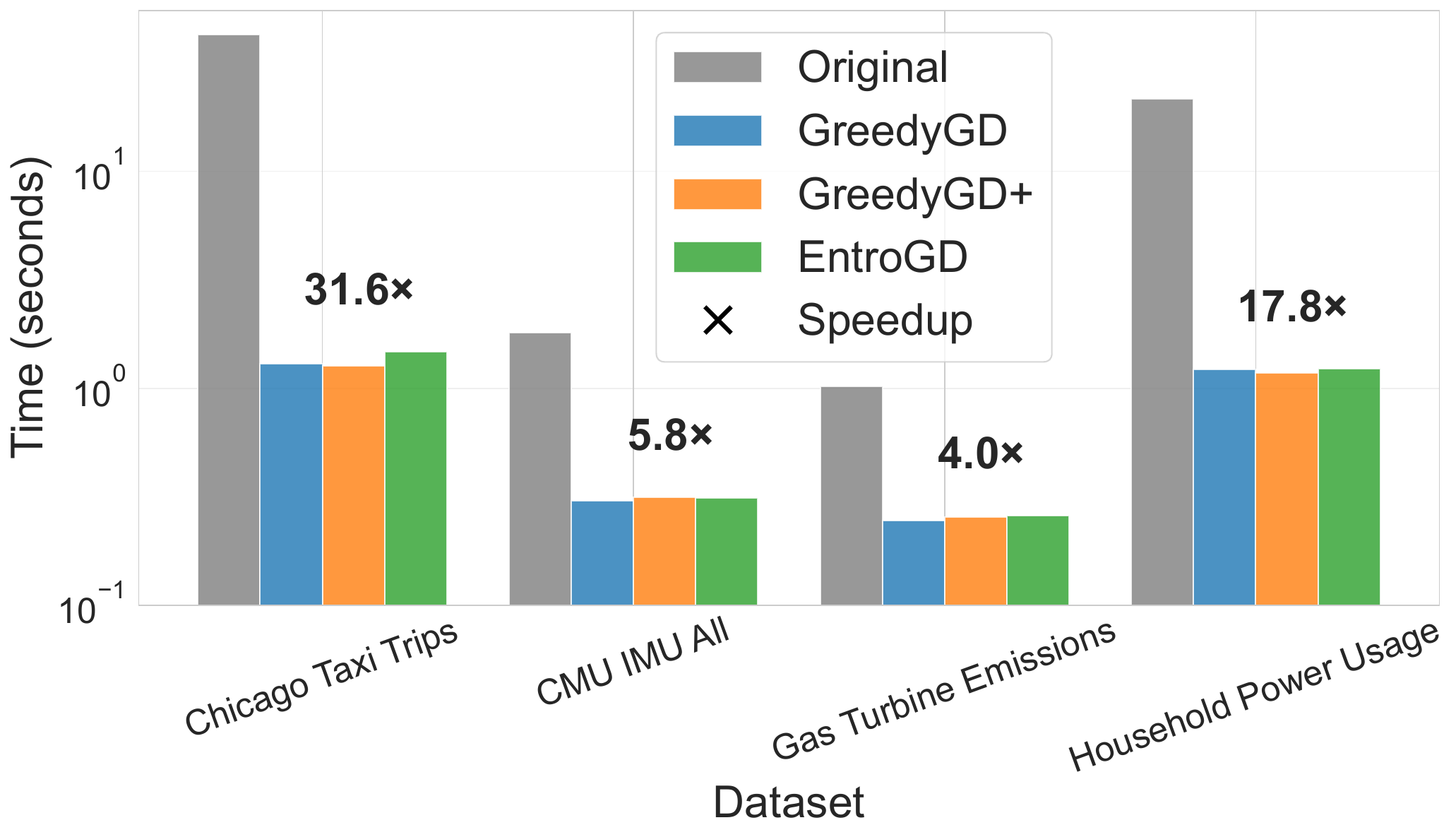}
    \caption{Clustering time comparison. 
    \vspace{-0.1in}
    }
    \label{fig:clustering_times}
\end{figure}

\subsection{Clustering Performance}
\label{sec:clustering_performance}

For each dataset, $k$-means clustering is first performed on the original data to obtain reference results, followed by clustering on the compressed representations from each GD-based method.
Clustering is repeated 10 times with 100 random initializations, and the Silhouette coefficient is computed on a random sample of $n = 10{,}000$ points to limit runtime.

\subsubsection{Clustering Quality}
\label{sec:clustering_quality}

As shown in Fig.~\ref{fig:metrics_scatter_side_by_side} and~\ref{fig:metrics_scatter_side_by_side_plus}, and summarized in Table~\ref{tab:clustering_performance_summary}, both EntroGD and GreedyGD+ achieve clustering quality nearly identical to that of the original data, whereas GreedyGD shows higher error.
Specifically, EntroGD and GreedyGD+ attain a median AR of 1.001 (1.009 for GreedyGD), indicating minimal distortion in clustering structure, and exhibit strong consistency with the original results, with median AMI scores of 0.961 and 0.968, respectively, compared with 0.912 for GreedyGD.
The Silhouette coefficients of EntroGD (0.393) and GreedyGD+ (0.394) are also higher than that of GreedyGD (0.389), indicating more cohesive and well-separated clusters.
The improvement stems from the storage and use of deviation means in analytics, which provide a more accurate summary of intra-base deviations compared to the centroid approximation used in GreedyGD.

\subsubsection{Clustering Time}
\label{sec:clustering_time}

Fig.~\ref{fig:clustering_times} shows that the GD-based methods yield similar clustering times since they use roughly the same number of condensed samples. They all achieve substantially faster clustering, up to 31.6$\times$ faster, than the original data, by operating on fewer samples with appropriate weighting.
Thus, all GD methods not only reduce storage requirements but also enable much faster analytics without (significantly) compromising clustering quality.

\begin{table}[t]
    \centering
    \caption{Performance Summary.}
    \label{tab:clustering_performance_summary}
    \begin{tabular}{lccccc}
        \toprule
        \textbf{Compressor} & \textbf{CR~$\downarrow$} & \textbf{ADR~$\downarrow$} & \textbf{AR~$\downarrow$} & \textbf{AMI~$\uparrow$} & \textbf{Silhouette~$\uparrow$} \\
        \midrule
        GreedyGD & 0.339 & \textbf{0.008} & 1.009 & 0.912 & 0.389 \\
        GreedyGD+ & 0.354 & 0.026 & 1.001 & \textbf{0.968} & \textbf{0.394} \\
        EntroGD   & \textbf{0.332} & 0.026 & \textbf{1.001} & 0.961 & 0.393 \\
        \bottomrule
    \end{tabular}
    \vspace{-0.15in}
\end{table}

\subsection{System Implications of ADR}
\label{sec:system_implications_of_adr}
Across all datasets, EntroGD achieves a median ADR of $2.6\%$, indicating that analytics can be performed by accessing only a small fraction of the original data.
Compared with GreedyGD, EntroGD incurs a modest increase in ADR due to the inclusion of condensed samples with deviation means, but this overhead remains negligible at the system level.
In practical IoT and cloud/edge deployments, such low ADR values translate into substantial reductions in data movement, memory access, and analytics processing cost.
As a result, EntroGD enables faster analytics execution and reduced resource consumption while maintaining high analytical fidelity.

\section{Conclusion}
\label{sec:conclusion}

This paper introduced EntroGD, an entropy-guided generalized deduplication framework that decouples analytics fidelity from compression optimization.
The results show that compact analytical summaries and entropy-based bit selection enable efficient analytics directly on compressed IoT data with low data access overhead.
By significantly reducing the amount of data required for analytics, EntroGD highlights the potential of compression-aware analytics as a scalable and system-efficient design approach for networked IoT data processing.

\bibliographystyle{IEEEtran}
\bibliography{ref}

\end{document}